\documentstyle[12pt]{article}

\begin{document}
\begin{center}
{\Large{ \bf Entangled probability distributions}}

 {\bf
Vladimir N. Chernega,$^1$ Olga V. Man'ko,$^{2,3}$ Vladimir I.
Man'ko$^{2,4}$}\\
\medskip
{\it
$^1$Russian University of Transport (MIIT), Moscow, Russia\\
\smallskip
$^2$Lebedev Physical Institute, Russian Academy of Sciences\\
Leninskii Prospect 53, Moscow 119991, Russia\\
\smallskip
$^3$Bauman Moscow State Technical University\\
The 2nd Baumanskaya Street 5, Moscow 105005, Russia\\
\smallskip
$^4$Moscow Institute of Physics and Technology (State University)\\
Institutskii per. 9, Dolgoprudny, Moscow Region 141700, Russia}\\
\end{center}
\smallskip
$^*$Corresponding author e-mail:~~~mankoov\,@lebedev.ru

\begin{abstract}\noindent
Concept of entangled probability distribution  of several random variables  is introduced.  These probability distributions describe multimode quantum states in probability representation of quantum mechanics. Example of entangled probability distribution is considered.
\end{abstract}
\medskip


\section{ Introduction.}
The foundations of conventional probability theory were considered long ago and are described , e.g. in books  \cite{Kolmogorov,Holevo,Randomness}. The probability theory is necessary to be used to describe random process \cite{Randomness} both classical and quantum. Also to  construct the theory of quantum phenomena like quantum mechanics, quantum statistics, quantum field theory one needs to apply all the aspects of the conventional probability theory.The aim of our work is to study the possibility to use the problem related to back action of the results obtained in quantum mechanics to clarify some aspects which are new in conventional probability theory. The point is that in quantum mechanics new interesting phenomenon called entanglement of quantum system states was found \cite{Sch1935,Sch36}. It is special types of correlations between the subsystems in multiparticle systems. For example, the review of the  entanglement phenomenon in information theory was discussed in \cite{1}. 
The phenomenon of entanglement is related to the properties of quantum systems like Bell inequalities \cite{Bell}. For example, quantum properties reflected from Bell inequalities were discussed in \cite{Khrennikov,AndreevMankoDubna,Andreev2007,Andreev2009,Entropy2022}.  On the other hand recently it was clarified that all quantum states of physical systems can be described by conventional probability distributions using the introduced probability representation of quantum mechanics  \cite{ManciniPL,ManciniFP,TombesiSemOpt, OVMVIMLasRes1997, Mancini2}. The spin tomography was introduced in \cite{Dodonov,JETP} and discussed in \cite{Bregenze,TerraCunha,Weigert,Paini}. 
In \cite{AndreevMankoDubna} in the frame of probability representation of  quantum mechanics the spin tomography was applied to Bell's inequality problem, and was shown that the  satisfaction or violation of Bell's inequalities can be understood as properties of tomographic functions for joint probability distributions for two spins. 

In the probability representation entanglement phenomenon is also associated with the specific properties of the quantum states which may be considered as properties determined by the specific properties of used probability distributions which we call entangled probability distributions. In our work we introduce autonomous notion of the entangled probability distribution and consider their properties on few examples induced by the existed entanglement phenomena in quantum mechanics like these phenomena in oscillator system states. 

In quantum mechanics very important phenomenon is the superposition principle. The superposition principle of wave functions \cite{SCH26} can be discussed using density matrices describing quantum states introduced by \cite{Landau,vonNeumann}. The nonlinear addition of density matrices describing quantum states corresponds to the linear combination of the wave function  \cite{MankoMarmoSudarshan,SudarshanJRLR,MankoMarmoSydarshanJRLR}.  We will use the map of tomograms and density matrices \cite{OVMVIMMarmoJPHYS,OVMVIMMarmoPHysScr,OVMVIMMarmoKrakow,OVMVIMMarmoVitale} to study properties  corresponding to entanglement phenomenon. The tomograms and the entanglement phenomenon in the two mode squeezed states and two-mode even and odd coherent states was considered in \cite{myspie,IzvRAN}.  The application of the symplectic tomography scheme to the  Stimulated Raman Scattering (SRS) and to the Stimulated Brillouin Scattering was done in \cite{Kuznetsov,SPIE54,VRMB,VRMB1} and the entanglement in these processes were discussed.  The classical propagator of SRS was introduced in \cite{1998}. The tomographic entropy, information, entanglement phenomenon and quantum distances between quantum states of photon--phonon mode of SRS in the probability representation were discussed in \cite{6256}. 

The quantum entanglement in the Raman Scattering was also investigated in \cite{Pathak,Perina}. It is necessary to add that optical tomograms of the time-evolved states generated by the evolution of different kinds of initial wave packets in a Kerr medium and optical tomograms of maximally entangled states generated at the output modes of a beam splitter was theoretically studied in \cite{Rohith}, \cite{Rohith1}. The entanglement criterion in probability representation was suggested in \cite{mySudarshan,heatingmap}.
In \cite{Facchi} a review of the Radon transform and the instability of the tomographic reconstruction process were discussed. 
In \cite{Polkovnikov} it was shown that in classical mechanics can be introduced  the Hermitian operators and the concepts of classical mechanics can be formulated using the inverse Wigner--Weyl transform of the classical probability distributions. In \cite{Khrennikov5} the attempts to study the problems of complex biosystems composed of many subsystems and in \cite{SocialLaser} the attempts to model social process using the language of quantum mechanics, thermodynamics, quantum information and field theory were done. In \cite{KhrennikovAlo} a review of classical probability representations of quantum states and observables is done and in \cite{Khrennikov1a} quantum nonlocality is discussed. New fundamental aspects of quantum mechanics based on groupoid approach are investigated in \cite{Marmo}. Development of the tomographic probability distribution method and different aspects of its applications are presented in \cite{Prud}. The classical and quantum aspects of tomography were discussed in \cite{FacchiLigabo63} and classical--quantum dynamics were considered in \cite{Elze}. The probability representation of quantum states is used recently to study properties of cosmology \cite{Stornaiolo62,Stornaiolo62a,M}. The review of the harmonic analyses of the density matrix properties including the method of symplectic tomography is given in \cite{Gouson,Gouson1,Gouson2}. The methods of star-product, tomography and probability representation of quantum mechanics were applied to different problems of quantum phenomena in   \cite{Foukzon,Belolipechkii,Bazrafkan,Filinov}. The geometrical interpretation of spin state in probability representation with the help of Malevich's squares is done in  \cite{Chernega1,Chernega2,Chernega3,Chernega4,Chernega5,Chernega6,Chernega7}.  

We give a review of the tomographic probability representation of quantum mechanics. In \cite{Chernega8} the formalism of quantum states and quantum observables using the formalism of standard probability distributions and classical-like random variables is presented and the evolution equation and energy spectra in the form of equations for probability distributions determining  the coherent and number states of photons are obtained.

Applications of quantum tomographic approach to different kinds of experiments as well as theoretical researches were discussed in \cite{Miroshnichenko,Koczor,23,12,Leon}.
The tomographic causal analysis of two-qubit states was done and theoretically was studied operations with a four-level superconducting circuit as a two-qubit system in \cite{Kiktenko1}. The tomograms of the state of a quantum circuit and of the Josephson junction were considered in \cite{myJos}.  The probability representation of the nonlinear coherent states of an ion in the Paul trap was introduced in \cite{Mynonlinear}, the Schr\"odinger cat states of a trapped ion in the Paul trap were discussed in \cite{cat}, the tomograms of an ion in the Paul and Penning traps were considered in \cite{3736}.  The tomograms of the states of the damped oscillator with Kronig--Penney excitation were considered in \cite{KronigPenneydamped}. The tomograms of the quark states were introduces in \cite{ActaPhysHung}. 

The consideration of classical mechanics within the framework of the tomographic representation was done in \cite{OVMVIMLasRes1997}, \cite{2004JRLR}, \cite{JRLR2001}, \cite{hlimit}, \cite{Pilyavets}. The relation between the quantum state description and the classical state description is elucidated in \cite{Ibort}. The quantum--classical limits for quantum tomograms are studied and compared with the corresponding classical tomograms in \cite{Marmo3}. The inverted oscillator was discussed recently in \cite{Schleih,12a} in connection with relation to cosmological problems.  The inverted oscillator in the probability representation of quantum mechanics was considered in \cite{myEntropy2023}. Quantum evolution equation for open system was considered in   \cite{TeretenkovNosal,Teretenkov,ManciniPaulieq}. 
Schr\"odinger and von Neumann equations, as well as equations for the evolution of open systems, are written in the form of linear classical–like equations for the probability distributions determining the quantum system states in \cite{Entropy2021}. 
The evolution of tomograms for different quantum systems, both finite and infinite dimensional was considered in \cite{Kishore}. A method based on tomographic representation to simulate the quantum dynamics was applied to the wave packet tunneling of one and two interacting particles in \cite{Arhipov}.

The paper is organized as follows. In Sec.2 we provide definition of entangled probability distributions and describe some their properties. 
In Sec.3 we present the examples of entangled probability distributions of two random continious variables. The conclusions and prospectives are given in Sec.4.  

\section{Probability distributions}
Let us consider the set of continious random variables $\vec x=(x_1,x_2,\ldots, x_N)$ and parameter $\vec a =(a_1,a_2,\ldots, a_N)$ which can take some values and we consider a function of these  variables $0\leq w(\vec x|\vec a)\leq 1
$. This function satisfies the condition 
\[\int w(\vec x|\vec a)\prod_{k=1}^N d x_k=1.\]
We consider this nonnegative normalized function as a conditional probability distribution of $N$ random variables. Then we can introduce nonnegative functions $w_k(x_k|a_k)$ using the definition 
\begin{equation}\label{eq.1}
w_k(x_k|a_k)=\int w(\vec x|\vec a)\prod_{k'\neq k=1}^N d x'_k.
\end{equation}
We call the probability distribution $w(\vec x|\vec a)$ separable one if it equals to convex sum of the products of functions (\ref{eq.1})
\begin{equation}\label{eq.2}
w(\vec x|\vec a)=\sum_s{\cal P}_s\prod_{s=1}^N w_s(x_s|a_s),\quad {\cal P}_s\geq0,\quad \sum_s{\cal P}_s=1.
\end{equation}
If the function $w(\vec x|\vec a)$ can not be represented in the form of convex sum of the functions (\ref{eq.1}) we call the probability distribution $w(\vec x|\vec a)$ entangled probability distribution, i.e. 
\begin{equation}\label{eq.3}
w(\vec x|\vec a)\neq\sum_s{\cal P}_s\prod_{s=1}^N w_s(x_s|a_s).
\end{equation}
In the case of two variables $x_1,x_2$ and parameters $a_1,a_2$ we have for entangled probability distribution $w(x_1,x_2|a_1,a_2)$ the definition 
\begin{equation}\label{eq.4}
w(x_1,x_2|a_1,a_2)\neq\sum_s{\cal P}_s w_s(x_1|a_1)w_s(x_2|a_2).
\end{equation}

\section{Examples of the entangled probability distributions}
Let us remind the value of the propagator for the inverted oscillator evolution 
\begin{equation}\label{propagator} G(x,y,t)=\frac{1}{\sqrt{2\pi\sinh t}}\exp\left[ \frac{i}{2}\left(\coth t(y^2+x^2)-\frac{2x y}{\sinh t}\right)\right].
\end{equation}
We will use this formula later. In order to consider examples we will study properties of two mode oscillator state. We consider the very simple model of state $\psi(x,y)$ of the form  
\begin{equation}\label{eq.2.1}
\psi(x,y)=\frac{1}{\sqrt2}\left(\psi_0(x)\psi_1(y)+\psi_1(x)\psi_0(y)\right)=\frac{x+y}{\sqrt\pi}\exp\left(-\frac{x^2}{2}-\frac{y^2}{2}\right).\end{equation}
The function (\ref{eq.2.1}) is the superposition of the wave functions of one-mode oscillators, where first function is ground state of the first oscillator
\begin{equation}\label{eq.2.2}
\psi_0(x)=\frac{e^{-\frac{x^2}{2}}}{\pi^{1/4}}
\end{equation}
and the second function is the first excited state of the second oscillator
\begin{equation}\label{eq.2.3}
\psi_1(y)=\frac{\sqrt2y}{\pi^{1/4}}e^{-\frac{y^2}{2}}.
\end{equation}
The normalization constant is such that we have
\begin{equation}\label{eq.2.4}
\int_{-\infty}^\infty\psi(x,y)\psi^\ast(x,y) d x d y=1.
\end{equation}
We calculated this normalization condition using the known formula 
\begin{equation}\label{eq.2.5}
\int_{-\infty}^\infty\exp\left(-a x^2+b x\right)d x=\sqrt{\frac{\pi}{a}}\exp\left(\frac{b^2}{4a}\right).
\end{equation}
Considering the derivative of both sides of the above equality with respect to the parameter $b$ we obtain the integral relation 
\begin{equation}\label{eq.2.6}
\int_{-\infty}^\infty x\exp\left(-a x^2+b x\right)d x=\sqrt{\frac{\pi}{a}}\frac{b}{2 a}\exp\left(\frac{b^2}{4a}\right).
\end{equation}
To calculate another integral we calculate derivative of the both sides of equality (\ref{eq.2.5}) with respect to the parameter $-a$ then we get the value of the integral   
\begin{equation}\label{eq.2.7}
\int_{-\infty}^\infty x^2\exp\left(-a x^2+b x\right)d x=\frac{1}{2a}\sqrt{\frac{\pi}{a}}\exp\left(\frac{b^2}{4a}\right)\left(1+\frac{b^2}{2a}\right).\end{equation}
Using the values of calculating integrals we get the normalization condition
\begin{equation}\label{eq.2.8}
\frac{1}{\pi}\int_{-\infty}^\infty(x+y)^2\exp\left(-x^2-y^2\right)d x d y=1.
\end{equation}
To consider the map of the wave function $\psi(x,y)$ onto the probability distribution let us study the integral
\begin{equation}\label{eq.2.10}
A=\int_{-\infty}^\infty\psi(x,y)\exp\left[i\left(\frac{\mu_1x^2}{2\nu_1}-\frac{X x}{\nu_1}+\frac{\mu_2y^2}{2\nu_2}-\frac{Y y}{\nu_2}\right)\right]d x d y.
\end{equation}
The integrals (\ref{eq.2.5})-(\ref{eq.2.7}) give the possibility to obtain the relation 
\begin{eqnarray}
&&A=\frac{\sqrt\pi}{\sqrt{\left(1-\frac{i\mu_1}{\nu_1}\right)\left(1-\frac{i\mu_2}{\nu_2}\right)}}
\left(-\frac{2i X}{\nu_1\left(1-\frac{i\mu_1}{\nu_1}\right)}-\frac{2i Y}{\nu_2\left(1-\frac{i\mu_2}{\nu_2}\right)}\right)\nonumber\\
&&\times\exp\left[-\frac{X^2}{2\nu_1^2\left(1-\frac{i\mu_1}{\nu_1}\right)}
-\frac{Y^2}{2\nu_2^2\left(1-\frac{i\mu_2}{\nu_2}\right)}\right].\label {eq.2.11}
\end{eqnarray}
The probability distribution related to these formulae reads (see, for example, \cite{Entropy2021}) 
\begin{equation}\label{eq.2.12}
w=\frac{A A^\ast}{4\pi^2|\nu_1||\nu_2|}.
\end{equation}
In explicit form the probability distribution $w(X,Y|\mu_1,\nu_1,\mu_2,\nu_2)$ is given by the relation 
\begin{eqnarray}
w(X,Y|\mu_1,\nu_1,\mu_2,\nu_2)=&&\frac{\left(\nu_2^2+\mu_2^2\right)X^2+2\left(\nu_1\nu_2+\mu_1\mu_2\right)X Y+\left(\nu_1^2+\mu_1^2\right) Y^2}{\pi \left(\nu_1^2+\mu_1^2\right)^{3/2}\left(\nu_2^2+\mu_2^2\right)^{3/2}}\nonumber\\
&&\times\exp\left[-\frac{X^2}{\mu_1^2+\nu_1^2}-\frac{Y^2}{\mu_2^2+\nu_2^2}\right].\label{eq.2.13}
\end{eqnarray}
For particular case $\nu_1=\nu_2=1,\quad \mu_1=\mu_2=0,$ one gets 
\begin{equation}\label{eq.2.14} w(X,Y|\mu_1=0,\nu_1=1,\mu_2=0,\nu_2=1)=\frac{1}{\pi}(X+Y)^2\exp\left(-X^2-Y^2\right). 
\end{equation}
One can check, that the function $w(X,Y,|\mu_1,\nu_1,\mu_2,\nu_2)$ (\ref{eq.2.14}) satisfy the condition 
\begin{equation}\label{eq.2.15}
\int\int w(X,Y,|\mu_1,\nu_1,\mu_2,\nu_2) d X d Y=1.
\end{equation}
As we know, this probability distribution function determines the quantum states which are entangled states. Due to this we call this probability distribution entangled probability distribution. In quantum mechanics the wave functions of two--mode oscillators which are obtained by means of superposition of two different wave functions are entangled pure states. In connection with this the tomographic probability distribution is described by the probability distribution function (\ref{eq.2.13}) and it cannot be represented in the form of equation (\ref{eq.2}). On the other hand the integral 
\begin{equation}\label{eq.2.16}
w(X|\mu_1,\nu_1)=\int w(X,Y|\mu_1,\nu_1,\mu_2,\nu_2)d Y=\frac{\exp\left(-\frac{X^2}{\mu_1^2+\nu_1^2}\right)}{\sqrt{\pi\left(\mu_1^2+\nu_1^2\right)}}
\left[\frac{1}{2}+\frac{X^2}{\mu_1^2+\nu_1^2}
\right]. 
\end{equation}
One can check, that 
\begin{equation}\label{eq.2.17}
\int w(X|\mu_1,\nu_1)d X=1.
\end{equation}
This function $w(X|\mu_1,\nu_1)$ is marginal conditional probability distribution of position $X$ which is the position of the first oscillator and the conditions are labeled by the real parameters $\mu_1$, $\nu_1$. Also, if we repeat analogous calculations for the second oscillator we will get  
\begin{equation}\label{eq.2.18}
w(Y|\mu_2,\nu_2)=\int w(X,Y|\mu_1,\nu_1,\mu_2,\nu_2)d X=\frac{\exp\left(-\frac{Y^2}{\mu_2^2+\nu_2^2}\right)}{\sqrt{\pi\left(\mu_2^2+\nu_2^2\right)}}
\left[\frac{1}{2}+\frac{Y^2}{\mu_2^2+\nu_2^2}.
\right]. 
\end{equation}
One can check, that 
\begin{equation}\label{eq.2.19}
\int w(Y|\mu_2,\nu_2)d Y=1.
\end{equation}
This function $w(Y|\mu_2,\nu_2)$ is marginal conditional probability distribution of position $Y$ which is the position of the second  oscillator and the conditions are labeled by the real parameters $\mu_2$, $\nu_2$.

Let us calculate other probability distributions given by two density operators ${\hat{\bar\rho}}_0$, $
{\hat{\widetilde 
\rho}}_1$ 
\begin{equation}\label{eq.2.20}
\hat\rho(0,1)=\frac{1}{2}\left({\hat{\bar\rho}}_0 {\hat{\widetilde 
		\rho}}_1+{\hat{\bar\rho}}_1 {\hat{\widetilde 
		\rho}}_0\right), 
\end{equation}
here ${\hat{\bar\rho}}_0(x,x')=\psi_0( x)\psi_0^\ast(x')$ and 
${\hat{\widetilde\rho}}_1(y,y')=\psi_1(y)\psi^\ast_1(y') $. This   density operator $\hat \rho(0,1)$ describes the separable states of the two-mode oscillator. Other two separable states are described by operators ${\hat{\bar\rho}}_0 {\hat{\widetilde 
	\rho}}_1$ and ${\hat{\bar\rho}}_1 {\hat{\widetilde 
\rho}}_0$. The state with probability distribution (\ref{eq.2.13}) 
is the entangled state. The three states which are separable states are described by the probability distributions 
\begin{eqnarray}\label{last1}
w_{\hat \rho(0,1)}(X,Y|\mu_1,\nu_1,\mu_2,\nu_2)
=\frac{\exp\left(-\frac{X^2}{\mu_1^2+\nu_1^2}-\frac{Y^2}{\mu_2^2+\nu_2^2}\right)}{\pi\left(\mu_1^2+\nu_1^2\right)^{1/2}\left(\mu_2^2+\nu_2^2\right)^{1/2}}\left(\frac{Y^2}{\mu_2^2+\nu_2^2}+\frac{X^2}{\mu_1^2+\nu_1^2}\right),
\end{eqnarray} 
\begin{equation}\label{last2}
w_{{\hat{\bar\rho}}_0 {\hat{\widetilde\rho}}_1}(X,Y|\mu_1,\nu_1,\mu_2,\nu_2)=
\frac{2Y^2}{\pi \left(\mu_1^2+\nu_1^2\right)^{1/2}\left(\mu_2^2+\nu_2^2\right)^{3/2}}
\exp \left(-\frac{X^2}{\mu_1^2+\nu_1^2}-\frac{Y^2}{\mu_2^2+\nu_2^2}\right),
\end{equation}
\begin{equation}\label{last3}
w_{{\hat{\bar\rho}}_1 {\hat{\widetilde 
		\rho}}_0}(X,Y|\mu_1,\nu_1,\mu_2,\nu_2)=\frac{2X^2}{\pi \left(\mu_2^2+\nu_2^2\right)^{1/2}\left(\mu_1^2+\nu_1^2\right)^{3/2}}
	\exp \left(-\frac{X^2}{\mu_1^2+\nu_1^2}-\frac{Y^2}{\mu_2^2+\nu_2^2}\right).
\end{equation}
Comparing (\ref{eq.2.13}) with the (\ref{last1}), (\ref{last2}), {\ref{last3}) we see that the entangled state probability distribution (\ref{eq.2.13})  is essentially different from the separable state probability distributions (\ref{last1}), (\ref{last2}), {\ref{last3}). 
		
		Let us calculate the marginal distributions for the probability distributions given by the formulae 
	(\ref{last1}), (\ref{last2}), ({\ref{last3}). We get  
		\begin{eqnarray}
		&&w_{\hat \rho(0,1)}(X|\mu_1,\nu_1)=\int w_{\hat \rho(0,1)}(X,Y|\mu_1,\nu_1,\mu_2,\nu_2) d Y=\nonumber\\
		&&\frac{1}{2\sqrt{\pi\left(\nu_1^2+\mu_1^2\right)}}
		\exp\left(-\frac{X^2}{\mu_1^2+\nu_1^2}\right)\left(1+\frac{2X^2}{\mu_1^2+\nu_1^2}\right), \label{last4} 
		\end{eqnarray}	
			\begin{eqnarray}
		&&w_{\hat \rho(0,1)}(Y|\mu_2,\nu_2)=\int w_{\hat \rho(0,1)}(X,Y|\mu_1,\nu_1,\mu_2,\nu_2) d X=\nonumber\\
		&&\frac{1}{2\sqrt{\pi\left(\nu_2^2+\mu_2^2\right)}}
		\exp\left(-\frac{Y^2}{\mu_2^2+\nu_2^2}\right)\left(1+\frac{2Y^2}{\mu_2^2+\nu_2^2}\right), \label{last5} 
		\end{eqnarray}	
			\begin{eqnarray}
		&&
		w_{{\hat{\bar\rho}}_0 {\hat{\widetilde\rho}}_1}(X|\mu_1,\nu_1)=\int w_{{\hat{\bar\rho}}_0 {\hat{\widetilde\rho}}_1}(X,Y|\mu_1,\nu_1,\mu_2,\nu_2) d Y=\nonumber\\
	&&
		\frac{1}{\sqrt{\pi\left(\nu_1^2+\mu_1^2\right)}}
		\exp\left(-\frac{X^2}{\mu_1^2+\nu_1^2}\right), \label{last6} 
		\end{eqnarray}	
		
			\begin{eqnarray}
		&&
		w_{{\hat{\bar\rho}}_0 {\hat{\widetilde\rho}}_1}(Y|\mu_2,\nu_2)=\int w_{{\hat{\bar\rho}}_0 {\hat{\widetilde\rho}}_1}(X,Y|\mu_1,\nu_1,\mu_2,\nu_2) d X=\nonumber\\
		&&
	\frac{2Y^2}{\sqrt{\pi}\left(\nu_2^2+\mu_2^2\right)^{3/2}}
		\exp\left(-\frac{Y^2}{\mu_2^2+\nu_2^2}\right),
		\label{last7} 
		\end{eqnarray}	
		
			\begin{eqnarray}
	&&
	w_{{\hat{\bar\rho}}_1 {\hat{\widetilde 
				\rho}}_0}(Y|\mu_2,\nu_2)=\int w_{{\hat{\bar\rho}}_1 {\hat{\widetilde 
				\rho}}_0}(X,Y|\mu_1,\nu_1,\mu_2,\nu_2) d X=\nonumber\\
	&&
	\frac{1}{\sqrt{\pi\left(\nu_2^2+\mu_2^2\right)}}
	\exp\left(-\frac{Y^2}{\mu_2^2+\nu_2^2}\right), \label{last8} 
	\end{eqnarray}	
	
	\begin{eqnarray}
	&&
	w_{{\hat{\bar\rho}}_1 {\hat{\widetilde 
				\rho}}_0}
	(X|\mu_1,\nu_1)=\int w_{{\hat{\bar\rho}}_1 {\hat{\widetilde 
				\rho}}_0}(X,Y|\mu_1,\nu_1,\mu_2,\nu_2) d Y=\nonumber\\
	&&
	\frac{2X^2}{\sqrt{\pi}\left(\nu_1^2+\mu_1^2\right)^{3/2}}
	\exp\left(-\frac{X^2}{\mu_1^2+\nu_1^2}\right),
	\label{last9} 
	\end{eqnarray}		
Comparing (\ref{eq.2.16}) and (\ref{eq.2.18}) with  (\ref{last4})-(\ref{last9}) we also see the difference of marginal distributions in entangled state (\ref{eq.2.13}) and in separable states (\ref{last1}), (\ref{last2}), (\ref{last3}). There exists a central mass tomography of two--mode oscillator \cite{center,center1,center2}.  These oscillator states are described by the probability distributions of one random variable (the center of mass position). We will study entangled and separable states of the oscillator in future publications.

\section{Conclusion}	
We conclude mentioning the main ideas to be used on the bases of the probability representation of quantum mechanics. In fact this approach provides the possibility to formulate new aspects of probability theory and group representation theory which play important role in quantum mechanics foundations. As we know the Jordan–Schwinger map \cite{Jordan,schwinger} provides the possibility to map the properties of matrices like the $N\times N$ matrices $A$, $B$, $C$ with commutation relation $[A,B]=C$ onto three operators $\hat A$, $\hat B$, $\hat C$ acting in the Hilbert space of $N$-dimensional oscillator states, namely, $A\leftrightarrow \hat A=\sum_{i, k=1}^N A_{i k}\hat a^\dagger_i\hat a_k$,  $B\leftrightarrow \hat B=\sum_{i, k=1}^N B_{i k}\hat a^\dagger_i\hat a_k$, $C\leftrightarrow \hat C=\sum_{i, k=1}^N C_{i k}\hat a^\dagger_i\hat a_k$, where $[\hat a_i,\hat a^\dagger_k]=\delta_{i k}$ and $\hat a_i$, $\hat a^\dagger_k$ are the annihilation and creation boson operators of the $N$-mode oscillator. The commutator $[\hat A,\hat B]=\hat C$ of the oscillator operators repeats the commutator properties of the matrices $A$, $B$, $C$. For example, the spin Pauli matrices commutation relations are reproduced by the  two-dimensional oscillator bosonic commutation relations and the states of $s$-spin are mapped onto the excited energy states of the energy levels of the oscillator. We get the results that spin states can be invertibly mapped onto energy levels states of the two-mode oscillator. From this observation follows an idea that all the spin states can be mapped on the probability distributions describing the states of the two-mode oscillator (or the Hermitte polynomials  \cite{yesterday1,yesterday2,yesterday3}). This provides the possibility to map all the spin states (irreducible representations  of rotation $SU(2)$ group) onto transforms of the probability distributions which is new aspect of the group representation theory. If to apply this idea to $N$-dimensional oscillator properties we extend the result to other group representation theory ($SU(N)$ group and for other Lie groups). The $s$-spin states can be described as the multi--mode states associated with probability distributions of multidimensional Hermite  polynomials. The approach will be considered in future publications.

\end{document}